\documentclass{article}
\newcommand{\bfr}{\begin{flushright}}
\newcommand{\efr}{\end{flushright}}

\newcommand{\Tr}{{\rm Tr~}}
 
\begin{document}
\title{Spontaneous Compactification of Bimetric Theory
}
\author{Nahomi Kan\footnote{kan@gifu-nct.ac.jp}
\\
{\small
Gifu National College of Technology,
Motosu-shi, Gifu 501-0495, Japan
}
\bigskip
\\
Takuya Maki\footnote{maki@jwcpe.ac.jp}
\\
{\small
Japan Women's College of Physical Education,
Setagaya, Tokyo 157-8565, Japan
}
\smallskip
\\
and
\smallskip
\\
Kiyoshi Shiraishi\footnote{shiraish@yamaguchi-u.ac.jp}\\
{\small
Yamaguchi University,
Yamaguchi-shi, Yamaguchi 753--8512, Japan}
}
\date{\today
}
\maketitle
\begin{abstract}
We propose a model of bimetric gravity in which the mixing of metrics
naturally gives a mass to a graviton by the compactification with flux
of two gauge fields in extra dimensions. We assume that each metric
in the solution for the background geometry describes the
four-dimensional Minkowski spacetime with an
$S^2$ extra space, though the two radii of $S^2$ for two metrics take
different values in general.  The solution is derived by the effective
potential method in the presence of the magnetic fluxes on the extra
spheres. We find that a massive graviton is governed by the
Fierz-Pauli Lagrangian in the weak field limit and one massless
graviton left in four dimensions.
\end{abstract}



\section{Introduction
\label{sec1}}
In the last ten years or so, there have been a number of attempts to
understand the mysterious components in the universe, called as
dark matter and dark energy. One of the possibilities is to consider
the modifications (for reviews,
\cite{Clif,Capo}) of the Einstein gravity, which has been the
theoretical basis for analyzing the universe as a whole. We have become
aware of ignorance of the nature of gravitational interaction at large
distances and at very small distance. This fact inspires a search for
modifications of the  general relativity at large and small distances.
Recently, many authors
 study massive graviton theory (for reviews,
\cite{Hin,deR}) with much interest for understanding cosmology and other
aspects of gravity.
The progress to the consistent massive theory has been attained by
introducing another (to be non-dynamical) metric in the theory
\cite{RGT,HRS}. Besides them, the bigravity theory or bimetric theory of
gravity has a long history since 1970's \cite{fg,big}, which describes
massless and massive gravitons in general. In the massive theories, the
mass of the graviton is given by hand. The lack of theoretical
explanation of the origin of the mass scale is a shortcoming of the
generic massive gravity models. 

The mixing term of two metrics gives a mass to a graviton.
In the present paper, we consider a model in which the mixing
originates from the fluxes of gauge fields in extra dimensions. 

A model of multiple gauge fields with a kinetic mixing has been
considered three decades ago \cite{Holdom} and studies on possible
consequences in similar models have repeatedly appeared until recent
times
\cite{BJ,BJK,SSY}. A possible stringy origin of such a type of models
has also been investigated \cite{FSSY1,FSSY2}.  In their models, the
Lagrangian for two gauge fields is written as in the form:
\begin{equation}
{\cal
L}=-\frac{1}{4}\left({F_1}_{\mu\nu}{F_1}^{\mu\nu}+
{F_2}_{\mu\nu}{F_2}^{\mu\nu}+2\alpha{F_1}_{\mu\nu}{F_2}^{\mu\nu}\right)\,,
\end{equation}
where $\alpha$ is a {\it dimensionless} constant.
The models suggest existence of exotic particles and candidates of
dark matter. Being motivated partially by the work,
we come to an idea of using two gauge fields as well as two
metrics in the theory. The mixing term of two gauge kinetic part can
play a role of mixing of two metrics simultaneously.

We consider the simplest compactification in a six-dimensional model
with $U(1)$ gauge field strengths in the extra space.
This yields the non-derivative interaction of two metrics in the
vierbein formalism~\cite{N1,N2,refbiz,HR,Alexandrov}. In the present
paper, we analyze our model as the on-shell level, and quantum nature in
the model is not discussed here.

In the next section, we define our model. 
In Sec.~\ref{sec3}, after assumptions for the
metrics with compactification and magnetic fluxes in the extra space
are declared, compactification with four-dimensional Minkowski spacetime
is investigated by means of the effective potential. In
Sec.~\ref{sec4},  the effective four-dimensional Lagrangian for
gravitons in the weak field limit is derived. Finally, we
summarize our work and give remarks about the outlook in 
Sec.~\ref{summary}.

\section{The model
\label{sec2}}
Our model has two metrics $g_{MN}$ and
$f_{MN}$, and two $U(1)$ gauge fields ${A_g}_M$ and
${A_f}_M$, where $M, N$ run over $0, 1, 2, 3, 5$, and $6$. The field
strengths are defined usually as
${F_g}_{MN}=\partial_M {A_g}_N-\partial_N {A_g}_M$
and
${F_f}_{MN}=\partial_M {A_f}_N-\partial_N {A_f}_M$.
We consider the following action for six
dimensional spacetime expressed as
\begin{equation}
S=S_g[g,F_g]+S_f[f,F_f]+S_{int}[g,f,F_g,F_f]\,,
\end{equation}
where
\begin{equation}
S_g=\int
d^6x\sqrt{-g}\left[\frac{1}{2\kappa_g^2}R_g-\frac{1}{4}g^{MK}g^{NL}{F_g}_{MN}
{F_g}_{KL}-\Lambda_g\right]
\end{equation}
and
\begin{equation}
S_f=\int
d^6x\sqrt{-f}\left[\frac{1}{2\kappa_f^2}R_f-\frac{1}{4}f^{MK}f^{NL}{F_f}_{MN}
{F_f}_{KL}-\Lambda_f\right]\,.
\end{equation}
Here $R_g$ and $R_f$ are the Ricci scalars constructed from the metric
$g$ and $f$, respectively. The quantities denoted by $\kappa_g,\kappa_f,
\Lambda_g,
\Lambda_f$ are constants.

To analyze the preferred action $S_{int}$ including the mixing term,
we introduce two sechbeins $e_g$ and $e_f$ which satisfy
\begin{equation}
{e_g}_M^A\eta_{AB}{e_g}^B_N=g_{MN}\,,\qquad
{e_f}_M^A\eta_{AB}{e_f}^B_N=f_{MN}\,,
\end{equation}
where $\eta_{AB}=\eta^{AB}=diag.(-1,1,1,1,1,1)$.
Now we adopt the following action $S_{int}$, written with help of
antisymmetric symbols:
\begin{equation}
S_{int}=-\frac{\alpha}{96}\int
d^6x
\,\epsilon^{MNRSTL}\epsilon_{ABCDEF}
{e_g}^A_M{e_f}^B_N{e_g}^C_R{e_f}^D_S{F_g}_{TL}
{F_f}_{JK}{e_g}^{EJ}{e_f}^{FK}\,,
\label{Sint}
\end{equation}
where $\eta_{AB}{e_{g(f)}}^{AM}{e_{g(f)}}^B_N=\delta^M_N$.
The {\it dimensionless } coupling constant $\alpha$ satisfies
$|\alpha|<1$. Note that this term has two
reflection symmetries $e_g\leftrightarrow -e_g$  and $e_f\leftrightarrow
-e_f$ independently, and an exchange symmetry $e_g
\leftrightarrow e_f$.
Though this action seems a bit bizarre, the other term in $S_{g(f)}$
can also be rewritten using sechbeins and one can see
\begin{equation}
\sqrt{-g}(F_g)^2=\frac{1}{48}\epsilon^{MNRSTL}\epsilon_{ABCDEF}
{e_g}^A_M{e_g}^B_N{e_g}^C_R{e_g}^D_S{F_g}_{TL}
{F_g}_{JK}{e_g}^{EJ}{e_g}^{FK}\,,
\end{equation}
\begin{equation}
\sqrt{-g}=\det e_g=\frac{1}{720}\epsilon^{MNRSTL}\epsilon_{ABCDEF}
{e_g}^A_M{e_g}^B_N{e_g}^C_R{e_g}^D_S{e_g}^E_{T}{e_g}^F_{L}
\end{equation}
and
\begin{equation}
\sqrt{-g}R_g=\frac{1}{48}\epsilon^{MNRSTL}\epsilon_{ABCDEF}
{e_g}^A_M{e_g}^B_N{e_g}^C_R{e_g}^D_S{R_g}^{EF}{}_{TL}\,,
\end{equation}
where ${R_g}^{EF}{}_{TL}$ is identical to the
coefficient in the curvature two form as
$\Theta^{EF}=\frac{1}{2}{R}^{EF}{}_{TL}dx^T
\wedge dx^L$. Thus the action of interacting part (\ref{Sint})
can be said to have a preferred style. Note that if two metrics are
identical, $S_{int}$ becomes $-\frac{\alpha}{2}\int d^6x
\sqrt{-g}{F_g}_{MN}{F_f}^{MN}$.

\section{Compactification of the background geometry
\label{sec3}}
Compactification with the flux in the
Einstein-Maxwell theory  was investigated by Randjbar-Daemi, Salam and
Strathdee (RSS) more than three decades ago~\cite{RSS1}.
Therefore, existence of a similar solution for flux compactification
is expected in our model.

Now, we assume that each metric describes a direct product of
four-dimensional flat spacetime and an extra two sphere, $S^2$.
We assume that the two metrics have different scales as
\begin{equation}
g_{mn}dx^mdx^n={a^2}(d\theta^2+\sin^2\theta\,d\varphi^2)\,,\quad
f_{mn}dx^mdx^n={b^2}(d\theta^2+\sin^2\theta\,d\varphi^2)\,,
\end{equation}
where $\theta$ and $\varphi$ are the standard coordinates on
$S^2$, $m, n=5, 6$,  and $a$ and $b$ are the radii of spheres. Then the
Ricci tensors associated with two metrics are independently given by 
\begin{equation}
{R_g}_{mn}=\frac{1}{a^2}g_{mn}\,,\qquad
{R_f}_{mn}=\frac{1}{b^2}f_{mn}\,,
\end{equation}
where 
$R_{mn}$ denotes the Ricci tensor.

Then, we suppose that the constant `magnetic' flux penetrates the extra
sphere, just as in the RSS model~\cite{RSS1,RSS2}. Namely, we set
\begin{equation}
F_g=dA_g=-\frac{n_g}{2ea^2}a\,d\theta\wedge a\sin\theta d\varphi
\end{equation}
and
\begin{equation}
F_f=dA_f=-\frac{n_f}{2eb^2}b\,d\theta\wedge b\sin\theta d\varphi\,.
\end{equation}
Here the electric charge $e$ has been chosen a common value, for
simplicity.

We seek the background solution with the four-dimensional flat
spacetime. 
According to work of Wetterich~\cite{Wetterich}, we use the method of
the effective potential for a static solution, instead of solving the
equation of motion derived from the action directly. We now define a
potential corresponding to the action and ans\"atze as
\begin{eqnarray}
V(a,b)&=&
a^2\left(-\frac{1}{\kappa_g^2a^2}+\frac{n_g^2}{8e^2a^4}+\Lambda_g\right)
+b^2\left(-\frac{1}{\kappa_f^2b^2}+\frac{n_f^2}{8e^2b^4}+\Lambda_f\right)
\nonumber \\
&
&+{2\alpha}ab\left(
\frac{n_gn_f}{8e^2a^2b^2}\right)\,.
\end{eqnarray}
If we take new quantities $y\equiv ab$ and $x\equiv b/a$, the
potential  takes the form:
\begin{equation}
V(x,y)=-\frac{1}{\kappa_g^2}-\frac{1}{\kappa_f^2}+\frac{1}{8e^2y}
\left({n_g^2}{}x+\frac{n_f^2}{x}+{2\alpha}
{n_gn_f}{}
\right)+y\left(\frac{\Lambda_g}{x}+\Lambda_fx\right)\,.
\label{pot}
\end{equation}
Then, the equations of motion are
expected to be satisfied if~\cite{Wetterich}
\begin{equation}
\left.\frac{\partial V}{\partial 
x}\right|_{x=x_0,y=y_0}=\left.\frac{\partial  V}{\partial 
y}\right|_{x=x_0,y=y_0}=0
\label{yy}
\end{equation}
and
\begin{equation}
V(x_0,y_0)=0\,.
\label{zz}
\end{equation}
The equation (\ref{zz}) requires a vanishing four-dimensional
cosmological constant. To make the equations (\ref{yy}, \ref{zz})
simultaneously satisfied, we must tune the values of $\kappa_g$ and
$\kappa_f$ to be specific values.

Since the last two terms in (\ref{pot}) is positive, the minimum
value of $V$ when the value of $y$ moves turns out to be
\begin{equation}
V(x,y_0)=-\frac{1}{\kappa_g^2}-\frac{1}{\kappa_f^2}+
\frac{1}{\sqrt{2}e}\sqrt{
\left({n_g^2}{}x+\frac{n_f^2}{x}+{2\alpha}
{n_gn_f}{}
\right)\left(\frac{\Lambda_g}{x}+\Lambda_fx\right)}\,,
\end{equation}
with
\begin{equation}
y_0=
\frac{1}{2\sqrt{2}e}\sqrt{
\left({n_g^2}{}x+\frac{n_f^2}{x}+{2\alpha}
{n_gn_f}{}
\right)\left(\frac{\Lambda_g}{x}+\Lambda_fx\right)^{-1}}\,,
\end{equation}

In a special case for
$\Lambda_g=n_g^{-2}\lambda$ and $\Lambda_f=n_f^{-2}\lambda$,
the minimum of $V(x,y_0)$ is achieved at
$x_0=\left|\frac{n_f}{n_g}\right|$ and it takes the value
\begin{equation}
V(x_0,y_0)=-\frac{1}{\kappa_g^2}-\frac{1}{\kappa_f^2}+
\frac{\sqrt{2\lambda}}{e}\sqrt{1+\alpha\, {\rm sign}(n_gn_f)}\,,
\end{equation}
and then 
$y_0=\frac{|n_gn_f|}{2e}\sqrt{\frac{1+\alpha\,{\rm
sign}(n_gn_f)}{2\lambda}}$.
Finally, we tune the constants as
\begin{equation}
\frac{1}{\kappa_g^2}+\frac{1}{\kappa_f^2}=
\frac{\sqrt{2\lambda}}{e}\sqrt{1+\alpha\, {\rm sign}(n_gn_f)}\,.
\end{equation}
In the present naive approach, the geometry of background is
determined, while the individual values for
$\kappa_g$ and 
$\kappa_g$ cannot be specified. The stability condition for flat
four-dimensional spacetime is needed for this determination.
We will see the condition in the subsequent section.

Taking further simplification, such that
$\Lambda_g=\Lambda_f\equiv\Lambda$,
$\kappa_g=\kappa_f\equiv\kappa$ and $n_g=n_f\equiv n$, we find
$x_0=1$ and $y_0=\frac{n}{2e}\sqrt{\frac{1+\alpha}{2\Lambda}}$, or
equivalently, $a^2=b^2=\frac{n}{2e}\sqrt{\frac{1+\alpha}{2\Lambda}}$,
$\frac{1}{\kappa^2}=
\frac{n\sqrt{\Lambda/2}}{e}\sqrt{1+\alpha}$.

\section{The masses of gravitons
\label{sec4}}
In this section, we consider the dynamical graviton modes of the lowest
excitation on the background. Here we
do not discuss on the Kaluza-Klein excited modes.

Provided that the background geometry is one obtained in the previous
section, the four-dimensional action for gravitons can be written by
\begin{eqnarray}
S^{(4)}&=&4\pi\int
d^4x\left\{
\sqrt{-g^{(4)}}\left[\frac{a^2}{2\kappa_g^2}R^{(4)}_g
+\frac{1}{\kappa_g^2}
-\frac{n_g^2}{8e^2a^2}-\Lambda_ga^2\right]\right.
\nonumber \\&
&\qquad\quad+\sqrt{-f^{(4)}}\left[\frac{b^2}{2\kappa_g^2}R^{(4)}_f
+\frac{1}{\kappa_f^2} -\frac{n_f^2}{8e^2b^2}-\Lambda_fb^2\right]
\nonumber \\& &\qquad\quad\left.
-\frac{\alpha}{12}\epsilon^{\mu\nu\rho\sigma}\epsilon_{abcd}
{e_g}^a_\mu{e_f}^b_\nu{e_g}^c_\rho{e_f}^d_\sigma\frac{n_gn_f}{8e^2ab}
\right\}\,,
\end{eqnarray}
where
$\mu, \nu=0, 1, 2, 3$ and $a, b=0, 1, 2, 3$ and the superscript `$(4)$'
indicates the four dimensional quantities constructed from four
dimensional metrics. Hereafter, $e^a_\mu$ stands for vierbeins. 

In the weak field limit \cite{N1,N2,refbiz}, {\it i.e.} $e_g=\eta
+\frac{1}{2}h_g$,
$e_f=\eta +\frac{1}{2}h_f$,
we find
\begin{equation}
\sqrt{-g^{(4)}}=\det
e_g=1+\frac{1}{2}[h_g]+\frac{1}{8}[h_g]^2-\frac{1}{8}[h_g^2]+O(h^3)\,,
\end{equation}
with a similar expression for $\sqrt{-f^{(4)}}$, and
\begin{eqnarray}
& &\frac{1}{24}\epsilon^{\mu\nu\rho\sigma}\epsilon_{abcd}
{e_g}^a_\mu{e_f}^b_\nu{e_g}^c_\rho{e_f}^d_\sigma\nonumber \\& &
=1+\frac{1}{4}([h_g]+[h_f])+\frac{1}{48}([h_g]^2+4[h_g][h_f]+[h_f]^2)
\nonumber \\&&
\quad-\frac{1}{8}([h_g^2]+4[h_gh_f]+[h_f^2])+O(h^3)\,.
\end{eqnarray}
Here $\eta$ denotes the four dimensional flat metric, and $[A]\equiv{\rm
tr~}A$ for notational simplicity. It is known that the asymmetric part
of $h$ can be omitted~\cite{refbiz}.

Now we can write down non-derivative terms in the
four-dimensional action.  The constant term disappears due to
Eq.~(\ref{zz}) for the background metrics obtained in the previous
section. The appearance of the linear term in $[h_g]$, $[h_f]$
bring about the instability of the flat four-dimensional spacetime. 
The vanishing of coefficients of linear terms in the
four-dimensional action tells us
\begin{eqnarray}
&&
\frac{1}{\kappa_g^2}
=\frac{n_g^2x_0}{8e^2y_0}+\frac{\alpha
n_gn_f}{8e^2y_0}+\Lambda_g\frac{y_0}{x_0}\,,\label{cg}
\\&
&\frac{1}{\kappa_f^2}
=\frac{n_f^2}{8e^2x_0y_0}+\frac{\alpha
n_gn_f}{8e^2y_0}+\Lambda_fx_0y_0\,.\label{cf}
\end{eqnarray}
Since Eq.~(\ref{zz}) still holds, the sums of each hand side of two
equations (\ref{cg}) and (\ref{cf})
are equal and the values for
$\kappa_g$ and
$\kappa_f$ can be obtained consistently.
Incidentally, for the case of the RSS model, the single condition
of vanishing cosmological constant in four dimensions (\ref{zz}) implies
the stability of the flat four-dimensional spacetime, because there is
only one gravitional field. Actually, Eqs.~(\ref{cg}, \ref{cf}) turn
out to be just the constraints which come from the variations
of two lapse functions
${e_g}^0_0$ and
${e_f}^0_0$.

When all the equations on the metrics including (\ref{cg}) and
(\ref{cf}) are satisfied, non-derivative term in the four-dimensional
action becomes very simple as
\begin{equation}
\frac{\alpha n_gn_f}{96e^2y_0}(([h_g]-[h_f])^2-[(h_g-h_f)^2])\,.
\end{equation}

On the other hand, the kinetic terms for graviton fields come from
\begin{eqnarray}
\int d^4x\sqrt{-g^{(4)}}R_g^{(4)}&=&\int
d^4x \left[-\frac{1}{4}\partial_\rho {h_g}_{\mu\nu}\partial^\rho
{h_g}^{\mu\nu} +\frac{1}{2}\partial_\rho {h_g}^{\rho}_{\ \
\mu}\partial_\nu {h_g}^{\nu\mu}\right.\nonumber \\&&\left.
\qquad\quad-\frac{1}{2}\partial_\mu {h_g}^{\mu\nu}\partial_\nu
{h_g}+\frac{1}{4}\partial_\rho {h_g} 
\partial^\rho {h_g}+O(h^3)\right]
\end{eqnarray}
and a similar expression for $f$, where $h\equiv [h]$ for simplicity.

Therefore the Lagrangian for linearized graviton fields is written by
\begin{eqnarray}
& &
\frac{y_0}{2\kappa_g^2x_0}\left[-\frac{1}{4}\partial_\rho
{h_g}_{\mu\nu}\partial^\rho {h_g}^{\mu\nu} +\frac{1}{2}\partial_\rho
{h_g}^{\rho}_{\ \
\mu}\partial_\nu {h_g}^{\nu\mu}\right.\nonumber \\&&\left.
\qquad\quad\quad\quad-\frac{1}{2}\partial_\mu {h_g}^{\mu\nu}\partial_\nu
{h_g}+\frac{1}{4}\partial_\rho {h_g} 
\partial^\rho {h_g}\right]\nonumber\\
& &
+\frac{x_0y_0}{2\kappa_f^2}\left[-\frac{1}{4}\partial_\rho
{h_f}_{\mu\nu}\partial^\rho {h_f}^{\mu\nu} +\frac{1}{2}\partial_\rho
{h_f}^{\rho}_{\ \
\mu}\partial_\nu {h_f}^{\nu\mu}\right.\nonumber \\&&\left.
\qquad\quad\quad\quad-\frac{1}{2}\partial_\mu {h_f}^{\mu\nu}\partial_\nu
{h_f}+\frac{1}{4}\partial_\rho {h_f} 
\partial^\rho {h_f}\right]\nonumber \\
&&+\frac{\alpha
n_gn_f}{96e^2y_0}\left[(h_g-h_f)^2-({h_g}_{\mu\nu}-{h_f}_{\mu\nu})^2
\right]\nonumber \\&&
=-\frac{1}{2}\partial_\rho
{H_0}_{\mu\nu}\partial^\rho {H_0}^{\mu\nu} +\partial_\rho
{H_0}^{\rho}_{\ \
\mu}\partial_\nu {H_0}^{\nu\mu}-\partial_\mu {H_0}^{\mu\nu}\partial_\nu
{H_0}+\frac{1}{2}\partial_\rho {H_0} 
\partial^\rho {H_0}\nonumber\\
& &
-\frac{1}{2}\partial_\rho
{H_1}_{\mu\nu}\partial^\rho {H_1}^{\mu\nu} +\partial_\lambda
{H_1}^{\lambda}_{\ \
\mu}\partial_\nu {H_1}^{\nu\mu}-\partial_\mu {H_1}^{\mu\nu}\partial_\nu
{H_1}+\frac{1}{2}\partial_\rho {H_1} 
\partial^\rho {H_1}\nonumber \\
&&+\frac{1}{2}\frac{\alpha
n_gn_f}{12e^2y_0^2}\left(\kappa_g^2x_0+\frac{\kappa_f^2}{x_0}\right)\left(H_1^2-{H_1}_{\mu\nu}^2
\right)\,,
\end{eqnarray}
where
\begin{equation}
H_0\equiv\frac{\frac{\kappa_f}{\kappa_gx_0}h_g+
\frac{\kappa_gx_0}{\kappa_f}h_f}{\sqrt{\frac{4}{y_0}
\left(\kappa_g^2x_0+\frac{\kappa_f^2}{x_0}\right)}}\,,\quad
H_1\equiv\frac{h_g-h_f}{\sqrt{\frac{4}{y_0}
\left(\kappa_g^2x_0+\frac{\kappa_f^2}{x_0}\right)}}\,.
\end{equation}

The quadratic term of $H_1$ corresponds to the Fierz-Pauli
mass term \cite{FP}. Therefore we conclude that the present model with
the  previously-obtained background geometry contains one massless
graviton field $H_0$ and one massive graviton field $H_1$. The mass
squared $m^2$ of
$H_1$ is given by
\begin{equation}
m^2=\frac{\alpha
n_gn_f}{12e^2y_0^2}\left(\kappa_g^2x_0+\frac{\kappa_f^2}{x_0}\right)\,,
\end{equation}
if $\alpha n_gn_f$ is positive.

Now we examine the simple cases seen already in the last part of the
previous section.
In the special case for
$\Lambda_g=n_g^{-2}\lambda$ and $\Lambda_f=n_f^{-2}\lambda$,
we found
$x_0=\left|\frac{n_f}{n_g}\right|$ and
$y_0=\frac{|n_gn_f|}{2e}\sqrt{\frac{1+|\alpha|}{2\lambda}}$. The
gravitational constants should be chosen as 
\begin{equation}
\frac{1}{\kappa_g^2}=\frac{1}{\kappa_f^2}=
\frac{\sqrt{\lambda/2}}{e}\sqrt{1+|\alpha|}\,.
\end{equation}
In this case, we find that the mass of the massive graviton is
\begin{equation}
m^2=\frac{2\sqrt{2\lambda}\alpha e
}{3(1+|\alpha|)^{3/2}}\frac{n_g^2+n_f^2}{n_g^2n_f^2}\,.
\end{equation}

In the further simple case for
$\Lambda_g=\Lambda_f\equiv\Lambda$,
$\kappa_g=\kappa_f\equiv\kappa$ and $n_g=n_f\equiv n$, we found
$x_0=1$ and $y_0=\frac{n}{2e}\sqrt{\frac{1+\alpha}{2\Lambda}}$,
$\frac{1}{\kappa^2}=
\frac{n\sqrt{\Lambda/2}}{e}\sqrt{1+\alpha}$, and $0<\alpha<1$.
The mass of the massive graviton is
\begin{equation}
m^2=\frac{4\sqrt{2\Lambda}\alpha e
}{3n(1+\alpha)^{3/2}}\,.
\end{equation}
Note that since the ratio 
$(m^2/\frac{2}{a^2})=\frac{\alpha}{3(1+\alpha)}$ is always smaller than
one, the massive graviton is expected to be lighter than the first
Kaluza-Klein excited mode of massless graviton.

\section{Summary and outlook
\label{summary}}
In the present paper, we presented a model of bimetric theory in six
dimensions. We showed that the compactification with
fluxes in the extra space leads to the four-dimensional massive and
massless gravity. The relation of parameters which allows the
compactification has been obtained. We found the relation in the mass of
the massive graviton and the parameters in the model.

It is interesting to see that  two radii of the extra spheres in two
metrics can have different values in general.  This fact will be of more
importance if we consider the Kaluza-Klein towers of excitation of
gravitons as well as matter fields. The possible variety of mass
spectrum is worth studying in some phenomenological and cosmological
contexts. It is interesting to study the model as a quantum field
theory, because of complexity of the interaction among the infinite
Kaluza-Klein excited modes of gravitons, gauge fields, and matter
fields to be added in the `asymmetric compactification' for
$a\ne b$. It is also interesting to introduce a dilaton fields into
bimetric models as in six dimensional supergravity model \cite{SS,MN}.
We think that similar compactifications can be considered in such
models. However, we suspect that there are possiblities of removing the
tuning of couplings and even to eliminate the asymmetry of two spheres
in models with a dilatonic field.

The cosmological application of our model attracts much attention,
for massive gravity is expected to solve the riddle of cosmic
acceleration. Our simple model is suitable to study of classical
and quantum cosmology, since cosmological aspects of the RSS model have
been studied by Okada~\cite{Okada,Okada2},
Halliwell~\cite{Halli,Halli2}, and the succeeding authors.

Finally, we notify that it is straightforward to generalize
the present model to the model of multigravity \cite{HR,MG}.
In such a theory, one may expect to find a new hierarchical spectrum of
gravitons and the other fields in the theory. It is interesting to
study such a model, because complicated particle content may cause
interesting quantum effects as well as novel cosmological evolution.

\section*{Acknowledgement
}
This study is supported in part by the Grant-in-Aid of Nikaido
Research  Fund.



\begin{thebibliography}{99}
\bibitem{Clif} T.~Clifton, P.~G.~Ferreira, A.~Padilla and C.~Skordis, 
Phys. Rep. {\bf 513} (2012) 1.
\bibitem{Capo} S.~Capozziello and M. De Laurentis, 
Phys. Rep. {\bf 509} (2011) 167.
\bibitem{Hin} K. Hinterbichler, 
Rev. Mod. Phys. {\bf 84} (2012) 671.
\bibitem{deR} C. de Rham, arXiv:1401.4173.
\bibitem{RGT}C. de Rham, G. Gabadadze and A. J. Tolley, Phys. Rev. Lett.
{\bf 106} (2011) 231101.
\bibitem{HRS} S. F. Hassan, R. Rosen and A. Schmidt-May,
JHEP {\bf 1202} (2012) 026.
\bibitem{fg} 
C.~J.~Isham, A.~Salam and J.~Strathdee, Phys. Rev. {\bf D3} (1971) 867;
A.~Salam and J.~Strathdee, Phys. Rev. {\bf D16} (1977) 2668;
A.~Salam and J.~Strathdee, Phys. Lett. {\bf B67} (1977) 429;
C.~J.~Isham and D.~Storey, Phys. Rev. {\bf D18} (1978) 1047.
\bibitem{big} 
T.~Damour and I.~I.~Kogan, Phys. Rev. {\bf D66} (2002) 104024;
D.~Blas, C.~Deffayet and J.~Garriga, Class. Quant. Grav. {\bf 23} (2006) 1697;
D.~Blas, C.~Deffayet and J.~Garriga, Phys. Rev. {\bf D76} (2007) 104036;
D.~Blas, Int. J. Theor. Phys. {\bf 46} (2007) 2258;
Z.~Berezhiani,
D.~Comelli, F.~Nesti and L.~Pilo, Phys. Rev. Lett. {\bf 99} (2007)
131101;
N.~Rossi, Eur. Phys. J. {\bf ST163}
(2008) 291;
Z.~Berezhiani,
D.~Comelli, F.~Nesti and L.~Pilo, JHEP {\bf 0807}
(2008) 130;
Z.~Berezhiani, F.~Nesti, L.~Pilo and N.~Rossi, JHEP {\bf 0907}
(2009) 083;
Z.~Berezhiani, L.~Pilo and N.~Rossi, Eur. Phys. J. {\bf C70}
(2010) 305;
M.~Ba\~nados, A.~Gomberoff and M.~Pino,
Phys. Rev. {\bf D84} (2011) 104028.
\bibitem{Holdom} B.~Holdom,
Phys. Lett. {\bf B166} (1986) 196. 
\bibitem{BJ} F.~Br\"ummer and J.~Jaeckel,
Phys. Lett. {\bf B675} (2009) 360. 
\bibitem{BJK} F.~Br\"ummer, J.~Jaeckel and V.~V.~Khoze,
JHEP {\bf 0906} (2009) 037. 
\bibitem{SSY} G.~Shiu, P.~Soler and F.~Ye,
Phys. Rev. Lett. {\bf 110} (2013) 241304. 
\bibitem{FSSY1} W.-Z.~Feng, G.~Shiu, P.~Soler and F.~Ye,
arXiv:1401.5880. 
\bibitem{FSSY2} W.-Z.~Feng, G.~Shiu, P.~Soler and F.~Ye,
JHEP {\bf 1405} (2014) 065.
\bibitem{N1} S.~G.~Nibbelink, M.~Peloso and M.~Sexton, Eur. Phys. J.
{\bf C51} (2007) 741.
\bibitem{N2} S.~G.~Nibbelink and M.~Peloso, Class. Quant. Grav. {\bf
22} (2005) 1313.
\bibitem{refbiz} C.~Bizdadea et al., JHEP {\bf 02} (2005) 016;
C.~Bizdadea et al., Eur. Phys. J. {\bf C48} (2006) 265.
\bibitem{HR} K.~Hinterbichler and R.~A.~Rosen,
JHEP {\bf 1207} (2012) 047.
\bibitem{Alexandrov} S.~Alexandrov,
Gen. Relativ. Grav. {\bf 46} (2014) 1639.
\bibitem{RSS1}
S.~Randjbar-Daemi, A.~Salam and J.~Strathdee, Nucl. Phys. {\bf B214}
(1983) 491.
\bibitem{RSS2}
S.~Randjbar-Daemi, A.~Salvio and M.~Shaposhnikov, Nucl. Phys. {\bf B741}
(2006) 236.
\bibitem{Wetterich}
C.~Wetterich, Phys. Lett. {\bf B113} (1982) 377.
\bibitem{FP} M.~Fierz and W.~Pauli, Proc. Roy. Soc. Lond. {\bf A173}
(1939) 211.
\bibitem{SS}
A.~Salam and E.~Sezgin, Phys. Lett. {\bf B147} (1984) 147.
\bibitem{MN}
K.~Maeda and H.~Nishino, Phys. Lett. {\bf B158} (1985) 381.
\bibitem{Okada} Y.~Okada, Phys. Lett. {\bf B150} (1985) 103.
\bibitem{Okada2} Y.~Okada, Nucl. Phys. {\bf B264} (1986) 197.
\bibitem{Halli} J.~J.~Halliwell, Nucl. Phys. {\bf B266} (1986) 228.
\bibitem{Halli2} J.~J.~Halliwell, Nucl. Phys. {\bf B286} (1987) 729.
\bibitem{MG}   
N.~Arkani-Hamed, H.~Georgi and M.~D.~Schwartz,
Ann. Phys. {\bf 305} (2003) 96;
N.~Arkani-Hamed and M.~D.~Schwartz,
Phys. Rev. {\bf D69} (2004) 104001;
M.~D.~Schwartz,
Phys. Rev. {\bf D68} (2003) 024029;
G.~Cognola, E.~Elizalde, S.~Nojiri, S.~D.~Odintsov and S.~Zerbini,
Mod. Phys. Lett. {\bf A19} (2004) 1435;
S.~Nojiri and  S.~D.~Odintsov,
Phys. Lett. {\bf B590} (2004) 295;
F.~Bauer, T.~Hallgren and G.~Seidl,
Nucl. Phys. {\bf B781} (2007) 32;
T.~Hanada, K.~Kobayashi, K.~Shinoda and K.~Shiraishi,
Class. Quant. Grav. {\bf 27} (2010) 225010;
S.~A.~Duplij and A. T. Kotvytskiy, Theor. Math. Phys. {\bf 177} (2013)
1400.
N.~Tamanini, E.~N.~Saridakis and T.~S.~Koivisto, 
JCAP {\bf 1402} (2014) 015;
C.~de Rham, A.~Matas and A.~J.~Tolley, 
Class. Quant. Grav. {\bf 31} (2014) 025004;
J.~Noller, J.~H.~C.~Scargill and P.~G.~Ferreira, 
JCAP {\bf 1402} (2014) 007.

\end{thebibliography}
\end{document}